# Quantum fluctuations in the open universe.

Ugo Moschella[1,2,3] and Richard Schaeffer[3]

[1] *Institut des Hautes Etudes Scientifiques, 91440 Bures-sur-Yvette*
[2] *Istituto di Scienze Matematiche Fisiche e Chimiche, Via Lucini 3, 22100 Como, and INFN sez. di Milano, Italy.*
[3] *Service de Physique Théorique, C.E. Saclay, 91191 Gif-sur-Yvette, France.*

We solve a continuing controversy when dealing with density fluctuations in open Friedman-Robertson-Walker (FRW) universes, on the physical relevance of a class of exponential modes. We show explicitly and rigorously that these modes enter the expansion of quantum fields. In the maximally symmetric de Sitter case, encountered in inflationary models, they are excited for fields with mass below a critical value $m_{cr}$. They are seen to be responsible for the breaking of the de Sitter symmetry for a massless field. We provide an exact calculation of the power spectrum for any mass. Our method is free of the divergences that appear in earlier treatments. We extend the construction to a generic open FRW universe.

It has been pointed out recently that the universe may have a lower-than-critical density ($\Omega < 1$) [1,2] and therefore a negative spatial curvature. This has considerably renewed the interest in spatially open FRW universes. Several authors have proposed possible scenarios in which an open universe may be realized in inflationary cosmology [2–5]. Moreover, there have been studies that explore the consequences of the assumed open structure of the universe [6,7]. The main challenge in both cases is the description of the density fluctuations to explain the observed large-scale structure of the universe. Quantum fluctuations are described by correlation functions $\mathcal{W}(x, x')$ between two events $x$ and $x'$. An open FRW universe admits $SO_0(1,3)$ as the symmetry group of the spatial sections and the evolution of the fluctuations at late times respects this symmetry. It is thus very important to separate space and time variables, in order to produce an expansion of such correlation functions that exhibits manifestly the desired symmetries. In principle, this can be done by using the modes of the spatial Laplacian. This problem turns out to be more difficult in the open case. The main source of controversy [6–9] stems out from the existence of *a continuum of unconventional modes*, which are of (real) exponential rather than oscillatory nature, and from the role they play in the description of quantum fluctuations. Despite the authors of [8] had the merit to discover that such modes may appear in calculations, due to uncontrolled statements there, confusion on this point persists. Indeed, astrophysical and cosmological computations which do not take into account the unconventional modes have continued appearing, while those calculations that have tried to include them phenomenologically run into difficulties due to the exponential divergences. The clarification of the mathematical and physical status of these modes is urgent, since they are potentially important at scales where the COBE measurements are relevant [10].

In this letter we introduce a method which for the first time provides unambigously the sought mode decomposition for a given correlation function. We first discuss the maximally symmetric open FRW universe, which can be identified with a region of a de Sitter (dS) space-time, and show explicitly and rigorously that the unconventional modes do enter the expansion of the two-point functions of Klein-Gordon (KG) fields for any value of the field mass $m$ less than the critical value $m_{cr} = \frac{1}{2R_v}\sqrt{d^2 - 2d}$, where $d$ is the space-time dimension and $R_v^{-1}$ is the curvature of the dS space-time. We then give the construction for a general open FRW universe.

Let us start by writing the metric of a $d$-dimensional FRW universe (it is useful to have the dimension $d$ as a free parameter): $ds^2 = dt^2 - a^2(t)dl^2$. With $(\frac{da}{adt})^2 = R_v^{-2} + (a\,R_s)^{-2}$, the curvature $R_v^{-1}$ and the comoving constant spatial curvature radius $R_s$ can be set equal to unity by a proper choice of time and length scales. One can then visualize the spatial section of the universe as the upper sheet $\Sigma$ of the $(d-1)$-dimensional hyperboloid with equation $\mathbf{x} \cdot \mathbf{x} = R_s^2 = 1$, embedded in a $d$-dimensional Minkowski ambient space with product $\mathbf{x} \cdot \mathbf{y} = \mathbf{x}^{(0)}\mathbf{y}^{(0)} - \mathbf{x}^{(1)}\mathbf{y}^{(1)} - \ldots - \mathbf{x}^{(d-1)}\mathbf{y}^{(d-1)}$ (there is no need to specify a particular set of coordinates on $\Sigma$). The spatial metric $dl^2$ is then that induced by the ambient Minkowskian metric. Consider now the Laplacian $\Delta$ on $\Sigma$ defined by the metric $dl^2$. The solutions of the equation $(\Delta + k^2)\psi(\mathbf{x}) = 0$ can be labelled by a complex number $q$ and a vector $\boldsymbol{\xi} = (\xi^{(0)}, \ldots \xi^{(d-1)})$ such that $k^2 = \left(\frac{d-2}{2}\right)^2 + q^2$, $\boldsymbol{\xi} \cdot \boldsymbol{\xi} = 0$ and $\xi^0 > 0$:

$$\psi_{iq}^{(d-1)}(\mathbf{x}, \boldsymbol{\xi}) = (\mathbf{x} \cdot \boldsymbol{\xi})^{-\frac{d-2}{2} + iq}; \qquad (1)$$

$k^2$ is real and non-negative both when $q$ is real and when $q$ is imaginary with $|q| \leq \frac{d-2}{2}$; $k$ may be thought of as the modulus of the wave number and $\boldsymbol{\xi}$ as the corresponding direction. The mode which is constant in space corresponds to $k^2 = 0$. The modes corresponding to the allowed imaginary values of $q$ are the unconventional modes we already mentioned. Since superpositions of them do not belong to the natural Hilbert space where the operator $\Delta$ is self-adjoint, most authors have been reluctant to use them in quantum theories on the space $\Sigma$. Still, one would like to know what their meaning is, whether they have to be included in the expansions of the correlation functions of (classical or quantum) fields and, in this case, to have a method to compute their weight.

Let us now restrict our attention to the dS space-time, which may be represented by a $d$-dimensional one-sheeted hyperboloid $X_d = \{x \in \mathbb{R}^{d+1} : x^{(0)2} - x^{(1)2} - \ldots - x^{(d)2} = -R_v^2 = -1\}$ embedded in a $(d+1)$-dimensional Minkowski ambient space-time whose scalar product is $x \cdot y = x^{(0)}y^{(0)} - x^{(1)}y^{(1)} - \ldots - x^{(d)}y^{(d)}$, $x^2 = x \cdot x$. The metric and the causal structure are induced on $X_d$ from the ambient space-time (see, e.g., [14]). Choosing the point $x_0 = (0, \ldots, 0, 1)$ as the "origin" of $X_d$, its future $\Gamma^+(x_0) = \{x \in X_d : (x - x_0)^2 > 0\}$ (the grey region in Fig. 1) is an open FRW universe with respect to the "cosmic time" $t = \operatorname{arc cosh} x^{(d)}$. With $a = \sinh t$, events of $\Gamma^+(x_0)$ have the coordinates

$$x^{(0,\ldots,d-1)} = \mathbf{x}^{(0,\ldots,d-1)} \sinh t, \ x^{(d)} = \cosh t, \quad \mathbf{x} \in \Sigma. \quad (2)$$

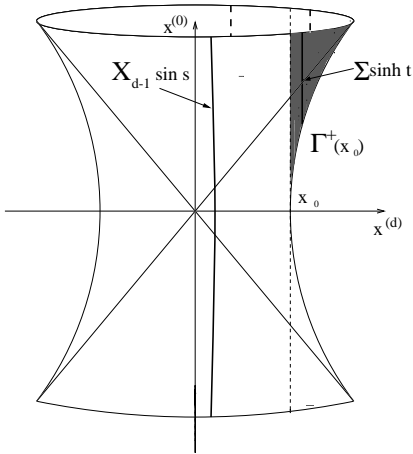

FIG. 1. de Sitter space-time

The dS-KG quantum fields can be described by their two-point Wightman functions $\mathcal{W}$. The latter are in this case already well known [11–14] and can be directly constructed in a manifestly dS invariant way [14] by exploiting suitable dS space-time plane waves solving the KG equation $(\Box + m^2)\psi(x) = 0$ ($\Box$ is the pseudo-Laplacian associated with the dS metric); these waves are similar to the spatial modes (1) but with the important difference that they are singular on three-dimensional light-like manifolds and are defined initially only on (suitable) halves of $X_d$. The physically relevant global waves are defined as boundary values from suitable domains of the complexified dS space-time $X_d^{(c)} = \{z \in \mathbb{C}^{d+1} : z^{(0)2} - z^{(1)2} - \ldots - z^{(d)2} = -1\}$ and namely from $\mathcal{T}^\pm = T^\pm \cap X_d^{(c)}$; $T^+$ and $T^-$ are the forward and backward tubes in the ambient complex Minkowski space-time $\mathbb{C}^{d+1}$ [13,14,16]. The dS plane waves are thus defined (and analytic) for $z \in \mathcal{T}^+$ or $z \in \mathcal{T}^-$ as

$$\psi_{i\nu}^{(d)}(z, \xi) = (z \cdot \xi)^{-\frac{d-1}{2} + i\nu}, \quad (3)$$

where $\nu \in \mathbb{C}$ and $\xi = (\xi^{(0)}, \ldots \xi^{(d)})$ are such that $m^2 = \left(\frac{d-1}{2}\right)^2 + \nu^2$, $\xi \cdot \xi = 0$ and $\xi^0 > 0$. These analyticity properties are equivalent to energy spectrum conditions. In particular, in the dS case, they give the Gibbons and Hawking temperature [12,14]. The waves (3) can be normalized in the KG [15] sense for all $m^2 > 0$. For $z \in \mathcal{T}^-$, $z' \in \mathcal{T}^+$ the Wightman function can then be represented as a superposition of plane waves [14]:

$$\mathcal{W}_\nu(z, z') = c_{d,\nu} \int \psi_{i\nu}^{(d)}(z, \xi) \psi_{-i\nu}^{(d)}(z', \xi) d\mu(\xi) \quad (4)$$

with $c_{d,\nu} = 2^{-(d+1)} \pi^{-d} \Gamma(\frac{d-1}{2} + i\nu) \Gamma(\frac{d-1}{2} - i\nu) e^{-\pi\nu}$. The integration is taken over all "directions" $\xi$ w.r.t. the appropriate measure $d\mu(\xi)$. Values of $\nu$ that are real or purely imaginary with $|\nu| < \left(\frac{d-1}{2}\right)$ correspond to physically acceptable Wightman functions. It follows from (4) that $\mathcal{W}_\nu(z, z') = C_{d,\nu} P^{(d+1)}_{-\frac{d-1}{2} - i\nu}(z \cdot z')$ where

$$P^{(d)}_{-\frac{d-2}{2} - i\nu}(u) = 2^{\frac{d-3}{2}} \Gamma\left(\frac{d-1}{2}\right) (u^2 - 1)^{\frac{3-d}{4}} P^{\frac{3-d}{2}}_{-\frac{1}{2} - i\nu}(u),$$
$$C_{d,\nu} = \frac{\Gamma(\frac{d-1}{2} + i\nu)\Gamma(\frac{d-1}{2} - i\nu)}{2^d \Gamma(\frac{d}{2}) \pi^{\frac{d}{2}}},$$

where $P_\rho^\mu$ ($Q_\rho^\mu$) are Legendre function [18]. $\mathcal{W}_\nu(z, z')$ can then be analytically continued into the "cut-domain" $\Delta = X_R^{(c)} \times X_R^{(c)} \setminus C$ where the cut $C = \{(z, z') \in X_d^{(c)} \times X_R^{(d)} : (z - z')^2 = \rho \geq 0\}$ reflects causality [13,14].

The problem we want to solve now amounts to expand the Wightman function (4) in terms of modes in which the time and space variables (2) are separated. The general solution of the KG equation with this property is given by $u_{a,b}(t, \mathbf{x}) = (\sinh t)^{\frac{2-d}{2}} \psi_{iq}^{(d-1)}(\mathbf{x}, \boldsymbol{\xi}) \left(a P_{i\nu - \frac{1}{2}}^{iq}(\cosh t) + b Q_{i\nu - \frac{1}{2}}^{iq}(\cosh t)\right)$, (labelled by $q$ and $\boldsymbol{\xi}$ as in Eq. (1), whereas the dS waves (3) are specified by $\xi$. One can now try to use canonical quantization to construct the Wightman functions (4) again, in terms of these modes. Following the prescriptions of canonical quantization [15] literally, one should therefore choose one particular $u_{\bar{a},\bar{b}}(t, x)$ to be the "positive frequency" mode. But here, strangely at first sight, one has to retain both the independent solutions [8]. The reason for this fact is that in the dS case the spatial manifold $\Sigma$ is not a complete Cauchy surface for the KG equation on $X_d$ (but is one half of such a surface). In a general open FRW universe one does not have access to information of this kind, which regards the global structure of the space-time manifold, and is lead to work only with the spatial manifold $\Sigma$ as if it were a Cauchy surface. Furthermore, this fact is also at the origin of the divergences one gets with unconventional modes.

To avoid all these oddities we follow an alternative method which bypasses them all. The result (see Eq. (8) and comments following) differs from the outcome of canonical quantization.

We saw that physically relevant Wightman functions are characterized by dS invariance and by precise analyt-icity properties in $X_d^{(c)}$. Knowledge of a correlation function with such properties within the region $\Gamma^+(x_0)$ implies knowledge of the latter everywhere in the dS space-time and we can consider its restriction to the manifold $X_{d-1}(s) \times X_{d-1}(s') = \{x, x' \in X_d, \ x^{(d)} = \cos s, \ x'^{(d)} = \cos s'\} \subset X_d \times X_d$ (see Fig.1):

$$\mathcal{W}_{s,s'}(\mathbf{x}, \mathbf{x}') = \mathcal{W}(x, x')|_{x^{(d)}=\cos s, \ x'^{(d)}=\cos s'} \qquad (5)$$

These points are obtainable from points in $\Gamma^+(x_0)$ by the replacement $t \to is$, $\mathbf{x} \to i\mathbf{x}$ in (2) through paths which are contained in the analyticity domain of $\mathcal{W}$. The two-point function $\mathcal{W}_{s,s'}$ is defined on a $(d-1)$-dimensional dS space-time $X_{d-1} = \{\mathbf{x} \in \mathbb{R}^d : \mathbf{x} \cdot \mathbf{x} = -1\}$. The relevant mode decomposition of $\mathcal{W}(x, x')$ can be obtained by finding a Källen-Lehmann decomposition for $\mathcal{W}_{s,s'}(\mathbf{x}, \mathbf{x}')$ [14] in terms of Wightman functions of KG fields on $X_{d-1}$. This can be done by computing the Laplace-type transform [14,17] of its retarded function $\mathcal{R}_{s,s'}(\mathbf{x}, \mathbf{x}')$ whose support is the causal cut of $\mathcal{W}_{s,s'}(\mathbf{x}, \mathbf{x}')$. Since $x \cdot x' = \sin s \sin s' \, \mathbf{x} \cdot \mathbf{x}' - \cos s \cos s'$, causality $(x-x')^2 = -2x \cdot x' - 2 \geq 0$ implies the following cut for $\mathcal{W}_{s,s'}(\mathbf{x}, \mathbf{x}')$: $\{\mathbf{x}, \mathbf{x}' \in X_{d-1} \times X_{d-1} : \mathbf{x} \cdot \mathbf{x}' = -\cosh v \leq -\cosh(\sigma - \sigma')\}$ where $\sinh \sigma = \cot s$ etc.. The sought transform is thus

$$G(q, s, s') = \frac{2\pi^{\frac{d-1}{2}}}{\Gamma(\frac{d-1}{2})} \int_{|\sigma-\sigma'|}^{\infty} Q^{(d)}_{-\frac{d-2}{2}-iq}(\cosh v) \times \qquad (6)$$
$$\times \mathcal{R}_{s,s'}(\cosh v) \sinh^{d-2} v \, dv$$

$$Q^{(d)}_{-\frac{d-2}{2}-iq}(z) = \frac{2^{\frac{d-3}{2}} e^{\frac{i\pi}{2}(d-3)} \Gamma(\frac{d-1}{2})(z^2-1)^{\frac{3-d}{4}}}{\pi} Q^{\frac{3-d}{2}}_{-\frac{1}{2}-iq}(z)$$

If $\mathcal{R}_{s,s'}(\cosh v)$ behaves as $e^{Kv}$ at infinity, $G(q, s, s')$ is analytic in the half-plane $\{q \in \mathbb{C} : \text{Im } q > K + \frac{d-2}{2}\}$. When $K = 0$, as for the cases we study in this letter, the inverse of the transform (6) is

$$\mathcal{W}_{s,s'}(\mathbf{x}, \mathbf{x}') = \frac{1}{2\pi i} \int_{-\infty+i\frac{d-2}{2}}^{+\infty+i\frac{d-2}{2}} G(q, s, s') \times \qquad (7)$$
$$\times C_{d-1,q} P^{(d)}_{-\frac{d-2}{2}+iq}(\mathbf{x} \cdot \mathbf{x}') 2q \, dq.$$

This expansion can then be analytically continued back to the physical region, where $\mathbf{x} \cdot \mathbf{x}' = \cosh r$ ($r \geq 0$). To apply this procedure to the Wightman functions of the KG field, an intermediate step to compute $G$ amounts to finding an integral representation for the relevant retarded function in a convolution form, since the transform (6) changes convolutions into products and we expect a factorized $G$. To this end it is crucial to use the integral representation (4). The result is:

$$G(q, s, s') = \frac{\gamma(iq)}{2} e^{-\pi q} \frac{P^{iq}_{-\frac{1}{2}+i\nu}(\cos s+i\epsilon)}{\sin^{\frac{d-2}{2}} s} \frac{P^{iq}_{-\frac{1}{2}-i\nu}(-\cos s'+i\epsilon)}{\sin^{\frac{d-2}{2}} s'}$$

where $\gamma(iq) = \Gamma\left(\frac{1}{2} - i\nu - iq\right) \Gamma\left(\frac{1}{2} + i\nu - iq\right)$. A major point in this formula is that all the singularities of $G(q)$ are the poles of the factor $\gamma(iq)$, the other terms being analytic in $q$. Using the analyticity properties of the r.h.s. of Eq.(7) we finally establish the key relation

$$\mathcal{W}_\nu(x, x') = \frac{1}{2\pi i} \int_{-\infty+i\frac{d-2}{2}}^{\infty+i\frac{d-2}{2}} C_{d-1,q} P^{(d)}_{-\frac{d-2}{2}+iq}(\cosh r) \times$$
$$\times \frac{P^{iq}_{-\frac{1}{2}+i\nu}(\cosh t) \, P^{iq}_{-\frac{1}{2}-i\nu}(-\cosh t'-i\epsilon)}{\sinh^{\frac{d-2}{2}} t \quad \sinh^{\frac{d-2}{2}} t'} \gamma(iq) e^{\pi q} q \, dq \qquad (8)$$

This result could have been obtained directly in the physical coordinates but the previous geometrical interpretation [14] would have been lost. For $d = 4$, Eq. (8) recalls the expansion given in [8] by canonical quantization but *is fundamentally different in that the integration over $q$ cannot in general be done on the real axis*. Note that this contour of integration, imposed by the large-distance behaviour of the two-point function, allows naturally to describe fields with all wave numbers down to $k = 0$ and consequently the spatially constant modes. When $m > m_{cr}$ there are no poles with Im $q > 0$ and by contour distortion the integration can be taken over the real $q$ axis. When $m < m_{cr}$ there are such poles in the Laplace transform and by contour distortion we get an expansion for $\mathcal{W}_\nu$ as an integral over the real $q$-axis plus discrete contributions from the poles of $\gamma(iq)$ at $0 < \text{Im } q < (d-2)/2$ (all lying on the imaginary $q$ axis). The number of such contributions is linked to the dimension of the space-time. When more generally $\mathcal{W}(x_1, x_2) = \int_0^\infty \rho(m) \mathcal{W}_{\nu(m)}(x_1, x_2) dm$, the expansion includes a continuum of exponential modes.

Using the more physical variable $k$ instead of $q$, we reduce Eq. (8) to an integration over the positive values of $k$ of quantities which are real and positive at $t = t'$:

$$\mathcal{W}_\nu(x, x') = (2\pi^2)^{-1} \sinh^{\frac{2-d}{2}} t \sinh^{\frac{2-d}{2}} t' \int_0^{+\infty} k \, dk \times$$
$$\times \left\{ \gamma_{osc}(k) \left[ e^{\pi q} P^{iq}_{-\frac{1}{2}+i\nu}(\cosh t) P^{-iq}_{-\frac{1}{2}-i\nu}(\cosh t') \right. \right.$$
$$\left. + e^{-\pi q} P^{iq}_{-\frac{1}{2}+i\nu}(-\cosh t + i\epsilon) P^{-iq}_{-\frac{1}{2}-i\nu}(-\cosh t' - i\epsilon) \right] \qquad (9)$$
$$+ \gamma_{exp}(k) P^{iq}_{-\frac{1}{2}+i\nu}(\cosh t) P^{iq}_{-\frac{1}{2}-i\nu}(\cosh t') \bigg\} \times$$
$$\times c_{d-1,q} \int d\mu(\boldsymbol{\xi}) \psi^{(d-1)}_{iq}(\mathbf{x}, \boldsymbol{\xi}) \psi^{(d-1)}_{-iq}(\mathbf{x}', \boldsymbol{\xi})$$

with $\gamma_{osc}(k) = \sinh \pi q \ \gamma(iq) \gamma(-iq) \ \theta(2k+2-d)$, $\gamma_{exp}(k) = \pi e^{\pi(q-\nu)} [\gamma(iq+i\epsilon) - \gamma(iq-i\epsilon)] \theta(d-2-2k)$ and $q = \sqrt{k^2 - \left(\frac{d-2}{2}\right)^2}$ (Im $q \geq 0$). We have written $C_{d-1,q} P^{(d)}_{-\frac{d-2}{2}+iq}(\cosh r)$ as a superposition of the spatial modes (1), similar to Eq. (4). Eq. (9) rather than Eq. (8) should be viewed as the sum over the modes labelled by the modulus of the momentum $k$ and the direction $\boldsymbol{\xi}$. The sum of two terms depending on time (for $k > 1$) is linked to the fact that $\Sigma$ is a half of a Cauchy surface. The spatial modes at $k < 1$ are not related by complex conjugation, but by the conjugation of the underlying Hilbert space product, fixed by the two-point function. These features are out of reach of the usual canonical quantization procedure. Let us examine the case $d = 4$ in more detail. For $m < \sqrt{2}$ there is exactly one pole contributing. The power spectrum $\mathcal{P}(k)$ is normalized as follows:

$$\mathcal{W}(x, x')|_{t=t'} = \int_0^\infty \frac{dk}{k} \mathcal{P}(k) P^{(4)}_{-1-iq}(\cosh r); \qquad (10)$$

here $k^2 = 1 + q^2$ and $P^{(4)}_{-1-iq}(\cosh r) = \frac{\sin qr}{q \sinh r}$. By defining $\mathcal{P}(k) = \mathcal{P}_{osc}(k) + \mathcal{P}_{exp}(k)$ we get

$$\mathcal{P}_{exp}(k) = \frac{\Gamma\left(\frac{3}{2}-\nu'\right)\Gamma(\nu')\left(\nu'-\frac{1}{2}\right)k\delta(k-k'_\nu)}{4\pi^3\sqrt{\pi}\sinh^{3-2\nu'} t}, \quad \mathcal{P}_{osc}(k) = \frac{qk^2\gamma_{osc}(k)\left(e^{\pi q}\left|P^{iq}_{-\frac{1}{2}+i\nu}(\cosh t)\right|^2 + e^{-\pi q}\left|P^{iq}_{-\frac{1}{2}+i\nu}(-\cosh t+i\epsilon)\right|^2\right)}{8\pi^3 \sinh \pi q \sinh^2 t}$$

with $\frac{1}{2} < \nu' = -i\nu < \frac{3}{2}$ and $k_{\nu'} = \sqrt{\frac{9}{4} - \nu'^2}$. For the massless case ($\nu' = \frac{3}{2}$) the discrete contribution is infinite. The corresponding $k = 0$ mode is thus seen to be responsible for the divergence of the Wightman function for dS invariant fields. In the context of inflationary models, the dS invariance is anyway only approximate, but nothing precludes the $SO_0(1,3)$ invariance of the spatial sections from holding at all times. The finite time at which inflation ends provides for a natural infrared cutoff in $k$. By subtracting the infinite constant we get the massless Wightman function, with $SO_0(1,3)$ symmetry.

To generalize the above approach to the case of a generic open FRW universe let us consider now an $SO_0(1, d-1)$ invariant two-point spatial correlation function $\Xi(\mathbf{x}, \mathbf{x}') = \Xi(u)$, defined for $x, x' \in \Sigma$, where $u = \cosh r = \mathbf{x} \cdot \mathbf{x}'$. Let us introduce the transform

$$F(q) = \frac{4\pi^{\frac{d-1}{2}}}{\Gamma\left(\frac{d-1}{2}\right)} \sin \pi\left(\frac{d-2}{2}+iq\right) \times \\ \times \int_1^\infty \Xi(u) Q^{(d)}_{-\frac{d-2}{2}-iq}(u)(u^2-1)^{\frac{d-3}{2}} du \quad (11)$$

$F(q)$ is analytic for $\mathrm{Im} q \geq \frac{d-2}{2} + K$ if $\Xi(u)$ is governed by $u^K$ at infinity. When $K = 0$ the inverse transform

$$\Xi(u) = \frac{1}{2\pi i} \int_{-\infty+i\frac{d-2}{2}}^{+\infty+i\frac{d-2}{2}} F(q) C_{d-1,q} P^{(d)}_{-\frac{d-2}{2}+iq}(u) 2q \, dq \quad (12)$$

can be obtained by means of the relation

$$\frac{4\pi^{\frac{d-1}{2}}}{\Gamma\left(\frac{d-1}{2}\right)} \sin \pi\left(\frac{d-2}{2}+iq\right) C_{d-1,q} \int_1^{+\infty} P^{(d)}_{-\frac{d-2}{2}+iq'}(u) \times \\ \times Q^{(d)}_{-\frac{d-2}{2}-iq}(u)(u^2-1)^{\frac{d-3}{2}} du = \frac{1}{q'^2-q^2} \quad (13)$$

In the special case where $F(q)$ has only real or purely imaginary singularities (i.e., only oscillatory or purely exponential modes), $\Xi(u) = \int_0^\infty \frac{dk}{k} \mathcal{P}(k) P^{(d)}_{-\frac{d-2}{2}+iq}(x)$. Conversely, if $\Xi(u)$ has the latter form, $F(q) = \int_0^\infty \frac{\mathcal{P}(k)}{q^2+(\frac{d-2}{2})^2-k^2} dk$ has the above singularities.

A two-point function $\mathcal{W}(x, x')$ on a generic open FRW universe defines at fixed times $(t, t')$ a two-point function $\Xi_{t,t'}(u) = \mathcal{W}(t, \mathbf{x}, t', \mathbf{x}')$ on $\Sigma$. We can then use the transform (11) to compute the associated spectrum $\mathcal{P}(k)$. In the dS case $\Xi_{t,t'}(u)$ is analytic in the complex $u$ plane except for the causal cut $(-\infty, \cosh(\tau-\tau'-i\epsilon))$ where again $\sinh \tau = \cot t$. By recasting Eq. (11) as an integral of the discontinuity of such an analytical $\Xi$ along its cut, we recover the Fourier-Laplace transforms of [17] recalled in Eqs. (6) and (7). Note that $F(q)$ and $G(q)$ may differ by an even function of $q$ with no singularity for $|\mathrm{Im} q| < \frac{d-2}{2}$ and nevertheless lead to the same function $\Xi(u)$. Two such functions also define the same power spectrum $\mathcal{P}(k)$, the latter being the physical quantity. If $\mathcal{W}(x, x')$ satisfies only $SO_0(1, d-1)$ invariance (neither analyticity nor dS invariance is required) the transform (11) provides the mode expansion of the correlation function in curved space. The motivation to introduce this transform and its relation to quantization are, however, fully justified by our treatment of the dS case.

We thank J.Bros, R.Balian and J.Y.Ollitrault for many discussions during the various stages of this work. We are especially indebted to K.Gorski for providing the original motivation to look into this problem. U.M. thanks the IHES and the SPhT for hospitality and financial support.

[1] P.J.E. Peebles, *Principle of Physical Cosmology*, (Princeton Univ. Press, 1993).
[2] B. Ratra and P.J.E. Peebles, Astrophys. J. **432**, L5 (1994). Phys. Rev. D **52**, 1837 (1995). B. Ratra Phys. Rev. D **50**, 5252 (1994).
[3] M. Bucher, A.S. Goldhaber and N. Turok, Nucl. Phys. **B**, Proc. Suppl. **43**, 173 (1995); Phys. Rev. D **52**, 3314 (1995).
[4] K. Yamamoto, M. Sasaki and T. Tanaka, Astrophys. J. **455**, 412 (1995).
[5] A.D. Linde, Phys. Lett. B **351**, 99 (1995). A.D. Linde and A. Mezhlumian, Phys. Rev. D **52**, 5538 (1995).
[6] D. Lyth and A. Woszczyna, Phys. Rev. D **52** 3338 (1995).
[7] A. Stebbins and R.R. Caldwell, Phys. Rev. D **52**, 3248 (1995).
[8] M. Sasaki, T. Tanaka, and K. Yamamoto, Phys. Rev. D**51**, 2979 (1995).
[9] M. Bucher and N. Turok, hep-th/9503393. Phys. Rev. D **52**, 5538 (1995).
[10] G.F. Smoot et al., Astrophys. J. **396**, L1 (1992).
[11] T.S. Bunch, P.C.W. Davies: Proc. R.Soc. Lond. A **360**, 117 (1978).
[12] G.W. Gibbons, S.W. Hawking: Phys. Rev. D **10**, 2378 (1977).
[13] J. Bros, J.-P. Gazeau and U. Moschella, Phys. Rev. Lett., **73**, 1746 (1994).
[14] J. Bros and U. Moschella, Rev. Math. Phys, **8**, 324 (1996). U. Moschella, Ann.Inst. Henri Poincaré **63**, 411-426 (1995).
[15] N.D. Birrell, P.C.W. Davies: Quantum fields in curved space, Cambridge University Press, Cambridge 1982.
[16] R.F. Streater, A.S. Wightman: PCT, Spin and Statistics, and all that W.A. Benjamin, New York 1964.
[17] J. Bros, G.A. Viano : Forum Mathematicum, **8**, 659 (1996).
[18] H. Bateman: Higher Transcendental Functions, Vol. I, Mc Graw-Hill, New York (1954).